\documentclass[twocolumn,prb,superscriptaddress]{revtex4}
\usepackage{graphicx}
\usepackage[dvipdfm]{hyperref}
\usepackage{dcolumn}

\begin{document}
\title{Entrapment of magnetic micro-crystals for on-chip electron spin resonance studies}

\author{N. Groll}
\affiliation{Department of Physics and the National High Magnetic Field Laboratory, Florida State University, 1800 E. Paul Dirac Drive, Tallahassee, Florida 32310, USA}

\author{S. Bertaina}
\affiliation{Department of Physics and the National High Magnetic Field Laboratory, Florida State University, 1800 E. Paul Dirac Drive, Tallahassee, Florida 32310, USA}
\affiliation{IM2NP-UMR6242 CNRS, Facult\'{e} des Sciences et Techniques,
Avenue Escadrille Normandie Niemen, 13397 Marseille Cedex 20, France}

\author{M. Pati}
\affiliation{Department of Chemistry and Biochemistry, Florida State University, Tallahassee, Florida 32306, USA}

\author{N.S. Dalal}
\affiliation{Department of Chemistry and Biochemistry, Florida State University, Tallahassee, Florida 32306, USA}
\affiliation{Department of Physics and the National High Magnetic Field Laboratory, Florida State University, 1800 E. Paul Dirac Drive, Tallahassee, Florida 32310, USA}

\author{I. Chiorescu}
\affiliation{Department of Physics and the National High Magnetic Field Laboratory, Florida State University, 1800 E. Paul Dirac Drive, Tallahassee, Florida 32310, USA}

\begin{abstract}
On-chip Electron Spin Resonance (ESR) of magnetic molecules requires the ability to precisely position nanosized samples in antinodes of the electro-magnetic field for maximal magnetic interaction. A method is developed to entrap micro-crystals containing spins in a well defined location on a substrate's surface. Traditional cavity ESR measurements are then performed on a mesoscopic crystal at 34 GHz. Polycrystalline diluted Cr$^{5+}$ spins were entrapped as well and measured while approaching the lower limit of the ESR sensitivity. This method suggests the feasibility of on-chip ESR measurements at dilution refrigerator temperatures by enabling the positioning of samples atop an on-chip superconducting cavity.
\end{abstract}

\date{Received 6 June 2009; accepted 20 July 2009. J. of Appl. Phys, DOI: 10.1063/1.3207774}
\maketitle

Research performed over the past years in the area of quantum information has made significant theoretical and experimental advances towards the realization of practical quantum computers. The basic operational unit, the qubit, is realized in two- or few-level quantum systems and several possibilities are currently being explored. Proposed qubits span the field of quantum physics from atoms in rare gases to solid-state systems, the later being the preferred candidates for building future quantum chips. In addition to a viable qubit, it is necessary to have a method to readout the quantum state of the qubit. Devices based on resonance techniques have gained significant momentum in the recent years (\textit{e.g.} Refs. \cite{Wallraff2004} and \cite{Bertet2004}) and have been applied to readout superconducting qubits. Such methods should be adjustable to sense spin qubits, or more generally, to study basic phenomena in quantum magnetism, as proven by first approaches based on SQUIDs \cite{Wernsdorfer2001} and Hall \cite{Wernsforfer2000,delBarco2003} techniques. 

For the purpose of spin-based quantum computing, molecular magnets are an attractive candidate due to their well-studied quantum properties \cite{Bertaina2008,Takahishi2009,Ardavan2007}. Synthesized in crystalline nanostructures, the magnetic molecules have a well defined size, shape and orientation allowing accurate characterization of their quantum mechanical parameters. Moreover, some molecules, with high spin ground state and high magnetic anisotropy, show tunable quantum phenomena via tunable tunneling channels through large zero-field energy barrier \cite{Wernsdorfer1999,Wernsdorfer2005,delBarco2003}. In terms of coherence times, other diluted spin systems have been proven to be even closer to the requirements of a quantum computer, as shown in systems like the nitrogen in C$_{60}$ (Ref. \cite{Morley2007}) and the nitrogen-vacancy (NV) color centers \cite{Hanson2008,Gurudev2007}. Additionally, recent measurements of coherence in magnetic ions diluted in a crystal matrix have demonstrated single and multiphoton Rabi oscillations at low \cite{Bertaina2007} and up to room temperature \cite{Nellutla2007,Bertaina2009}. 

An experimental challenge, particularly applicable to spin qubits due to their nanosized dimensions, is related to the controlled positioning of the spins in an area that can be readout by the output electronics. In this paper we address this issue, and introduce an experimental method allowing the placement of nanosized samples in well defined locations of a chip. In addition, to show that the spins environment is not altered by the chemical procedures, we perform traditional cavity ESR measurements on samples containing a sufficiently large amount of spins to be detectable by a 34 GHz Q-band setup. The entrapping method is applied to the well-known ESR calibration standard 1,3-bisdiphenylene-2 phenylallyl (BDPA) and to K$_3$NbO$_8$ nanocrystals containing diluted magnetic ions of Cr$^{5+}$(S=1/2) \cite{Nellutla2007}. Note that the method can be extended to molecular nanomagnets as well. In the case of diluted Cr$^{5+}$ spins in a K$_3$NbO$_8$ lattice, decoherence is due only to superhyperfine interactions in the neighboring K and Nb atoms, where there are no electronic spins \cite{Nellutla2007,Nellutla2008}. This diluted Cr system is also quasi-isotropic making it an ideal candidate for on-chip applications and an excellent demonstration for the method proposed here.

\begin{figure}
		\includegraphics[width=3.2 in]{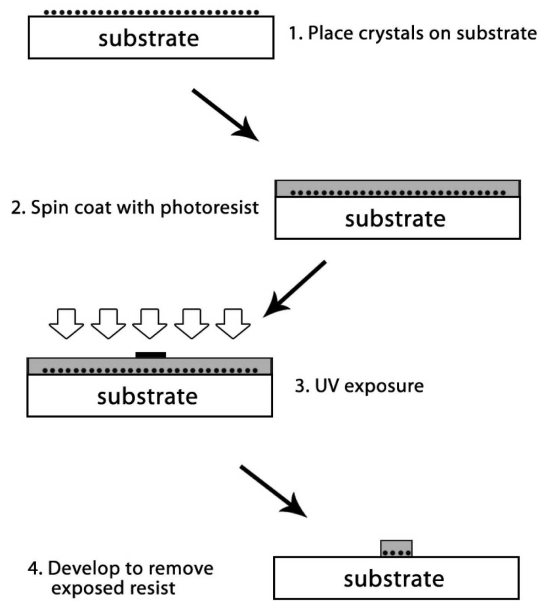}
	\caption{Graphic representation of the entrapment technique developed for microcrystal placement. The substrate is first coated with crystals (black dots) with precision tip tweezers and a brushing tip to obtain a uniform layer. Photoresist (shown by the grey boxes) is then placed on top of the crystals and spun at 5000 rpm for 30 seconds. Next, the sample is exposed to UV light wherever not blocked by the pattern mask (black rectangle). This defines the region where crystals are to remain. Finally, the sample is developed to remove the exposed regions.}
\label{fig1technique}
\end{figure}

Placement of magnetic microcrystals requires the ability to non-destructively deposit crystals without introducing contaminants that would potential interfere with the measurements. The method developed for microcrystal placement makes use of standard photolithography techniques as a means for defining the placement region (see Fig.~\ref{fig1technique}). Microcrystals are first ground into a powder then placed on a clean substrate surface using precision tip tweezers near the region where crystals are desired. It is important to stress that is necessary for the crystals to be in direct contact with the surface prior to applying photoresist because the crystals can act as a mask themselves. In such situations, undeveloped photoresist may act as an adhesion layer preventing the removal of the crystals outside the desired area. To create a uniform layer of powder, a fine tip is brushed over the crystals while viewing under a microscope.

Next a thick, positive photoresist (AZ4620) is placed on top of the crystals and spun at 5000 rpm's for 30 seconds. This resist was chosen for two reasons. First, the layer must be thick enough to completely encapsulate the crystals of micron scale thickness. We have observed that, in the case of micron sized crystals, other optical or electron-beam resist do not perform so well. Second, compared with other resists, this type of resist contains fewer solvents that could potentially be harmful to the crystals. In general, for an adequate choice of the resist, one should know as much as possible about the composition of the nanocrystal sample, and avoid choosing resists that could dissolve or alter its structure.  Following the spin coating process, the sample is placed immediately on a hot plate at 110$^{\circ}$C for 60 seconds to harden the resist and evaporate out any residual solvent. After the pre-exposure bake, the sample is cooled in a stream of nitrogen gas. Once cooled, the chip is exposed to ultra-violet light for about 4 seconds using a mask aligner, thus defining the pattern where the crystals are to be located. The sample is then immersed in the developer solution and agitated to remove the exposed photoresist regions (30-60 seconds). We noticed that in some cases it was necessary to place the sample in an ultrasonic bath during development in order to completely remove the crystals from the exposed region.

\begin{figure}
		\includegraphics[width=3.25in]{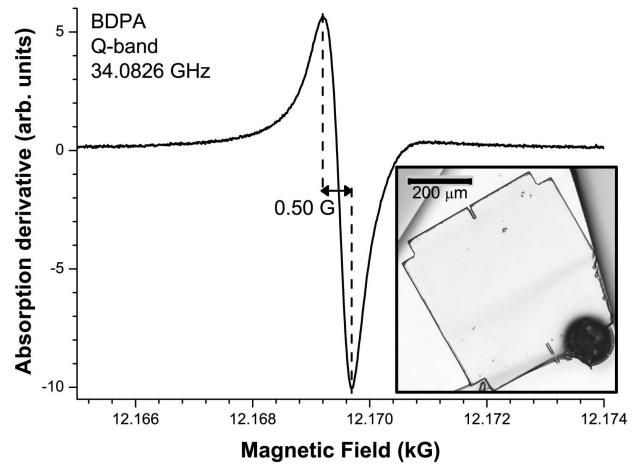}
	\caption{Room temperature ESR measurement of one BDPA crystal entrapped in optical photoresist. The resonance width is 0.50 G and the corresponding g-factor is 2.001.  The insert shows an optical microscope image of the measured crystal (dark circle in lower right corner of the designed square box) as placed on a quartz glass. The crystal is entrapped in photoresist (the visible 500 micron box with corners cut out).}
	\label{resBDPA}
\end{figure}

\begin{figure}
		\includegraphics[width=3.25in]{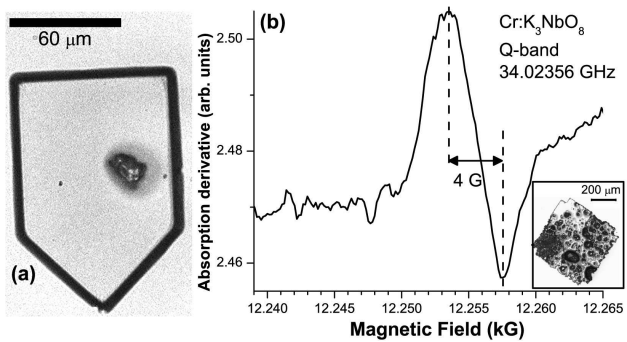}	
	\caption{(a) Photograph of a single crystal of Cr:K$_3$NbO$_8$ entrapped in the area delimited by the dark contour. The crystal is encased in optical photoresist, as shown by the grey surface of the contour. (b) Room temperature ESR measurement of diluted Cr$^{5+}$ spins. The insert shows the actual sample, consisting of few tens of single crystals, in order to achieve a measurable signal by the Q-band spectrometer. The obtained signal is the result of 40 averages, as required by the small size of the signal.}
	\label{resCr}
\end{figure}

The method presented here was successfully tested for two different types of materials, as described below. Following each application, testing of the integrity of the spin environment was performed by ESR characterization. The obtained samples were measured at room temperature using a Bruker Elexsys-500 EPR (Electron Paramagnetic Resonance) setup operating at 34 GHz and using a standard, TE$_{011}$ cylindrical cavity. Optimization of the measured signal-to-noise, microwave power and resolution was obtained by a proper adjustment of the field modulation, required to perform the derivative of the absorption curves shown in Figs.~\ref{resBDPA} and \ref{resCr}.

For a first application of the method, we entrapped a single crystal of BDPA on the surface of a quartz chip. BDPA has a known free radical concentration of 1.6$\times$10$^{19}$g$^{-1}$ and a Zeeman factor $g=2.0025$. \cite{Konovalov2003} The derivative of the absorption curve measured at $\sim 34$~GHz is shown in Fig.~\ref{resBDPA}. The signal is highly visible indicating a typical resonance width of 0.50~G=0.050~mT and a g-factor $g=hf/(\mu_BB)=2.001$ where $h$ is Planck's constant, $\mu_B$ the Bohr magneton and $f$ the resonant frequency (see Fig. \ref{resBDPA}). This data shows a non-destructive placement technique for a rather sensitive sample as the BDPA.

The same entrapping technique was used to position a single crystal of Cr$^{5+}$ spins $1/2$ sample highly diluted in a potassium niobate non-magnetic matrix (Cr:K$_3$NbO$_8$). The estimated concentration of spins is 0.03\% and the consequent number of spins in a single nanocrystal [see Fig.~\ref{resCr}(a)] is below the spin sensitivity of the ESR setup used here. In  order to perform the spin resonance study, another sample containing few tens of crystals was prepared [see insert of Fig.~\ref{resCr}(b)]. Due to the low spin concentration, we approached the limit of the measurement sensitivity, estimating the number of spins to be on the order of 10$^{12}$. 

The derivative of the absorption spectra shows a small resonance signal (see fig. \ref{resCr}) of width $\Delta B_{pp}=4$~G corresponding to lifetime $T_{2}^{*}=2\sqrt 2/\Delta f=250$~ns, with $\Delta f=f\Delta B_{pp}/B_{res}$; $f$ and $B_{res}$ are the resonant frequency and field. This value of $T_{2}^{*}$ agrees well with published data.\cite{Nellutla2007} The position of the resonance corresponds to a Zeeman factor $g=1.9835$, in agreement with the measured values for this quasi-isotropic crystal ($g_{\perp}=1.9878$, $g_{||}=1.9472$, from Refs. \cite{Nellutla2007} and \cite{Cage1997}). Note that in powder ESR, one generally observes a strong signal for $g_{\perp}$ and a much smaller one for $g_{||}$. In our case, the relatively large number of crystals explains the powder-like behavior of the signal, but the small number of spins makes detectable only the resonance corresponding to $g_{\perp}$. To insure that the measured signal is due to the Cr$^{5+}$ and not to the photoresist encasing it, a bare photoresist sample on quartz glass was measured and no interfering background signal was observed. 

The presented crystal placement method can be applied to entrap any type of magnetic nanocrystal on a chip, like the single molecule magnets (SMM) or diamond samples containing NV color centers. Such magnetic objects have a small magnetic moment making them inherently difficult to measure. Changes in spin orientation (i.e.~$+1/2$~$\rightarrow$~$-1/2$) will perturb only slightly the cavity and therefore a high quality factor cavity is required. The Cr$^{5+}$ spectrum shown in Fig.~\ref{resCr} actually shows that traditional cavity ESR has limited benefits when dealing with nanosized samples. On-chip ESR technique holds the promise of increased sensitivity, by using microstrip or superconducting cavities. Recent research has shown the ability to fabricate on-chip superconducting resonators with quality factors of 10$^5$-10$^6$ (Ref. \cite{Wallraff2004}) through the use of readily available standard photo or electron-beam lithography techniques. Their small size also allows the resonators to be easily incorporated with dilution refrigerators for millikelvin studies of spin-based qubits.

In conclusion this technique shows the ability to non-destructively place magnetic microcrystals in a well defined location. Further plans for scaling down the entrapment region are being conducted with the requirement of a more sensitive measurement technique. Superconducting resonators are a promising means for attaining additional sensitivity as well as performing on-chip ESR studies at millikelvin temperatures. This research is necessary for determining the viablity of micro/nano-crystals and potentially molecular magnets as qubits for quantum computation. 

We acknowledge support by the NSF Cooperative Agreement Grant No. DMR-0654118, NSF grants No. DMR-0645408, No. DMR-0506946, and No. DMR-0701462, the MARTECH institutre (FSU), the NHMFL (IHRP-5059), DARPA (HR0011-07-1-0031) and the Alfred P. Sloan Foundation.

\end{document}